\documentclass[11pt]{article}
\usepackage{graphicx}
\renewcommand{\baselinestretch}{1}
\begin{document}
\title{\bf  A fundamental requirement \\for crystal-field parametrization}
\author{\bf J. Mulak$^{1}$, M. Mulak$^{2}$}
\date{{\it  $^{1}$ Trzebiatowski Institute of Low Temperature
            and Structure Research,\\
            Polish Academy of Sciences, 50--950, PO Box 1410,
            Wroclaw, Poland\\
            $^{2}$ Institute of Physics,
            Wroclaw University of Technology,\\
            Wyb. Wyspianskiego 27,
            50--370 Wroclaw, Poland}}
\maketitle
\vspace*{0.2cm}
\noindent
{\bf Corresponding author:}\\
Prof. Jacek Mulak\\
Trzebiatowski Institute of Low Temperature and
Structure Research,\\
Polish Academy of Sciences,
50--950, PO Box 1410,
Wroclaw, POLAND\\
Tel: (+4871) 3435021, 3443206;  Fax: (+4871) 3441029\\
E-mail: Maciej.Mulak@pwr.wroc.pl
\renewcommand{\baselinestretch}{1.5}
\begin{abstract}
\noindent
The physically correct parametrization of the energy levels of transition ions in crystals in terms of
crystal-field (CF) Hamiltonians ${\cal H}_{\rm CF}=\sum_{k}\sum_{q}B_{kq}C_{q}^{(k)}$ has to be based on
the CF parameters $B_{kq}$ that lead to the correct CF splitting second moments, both the global one $\sigma$
and the partial ones $\sigma_{k}$.
Only such parametrizations correspond to the appropriate multipolar structure of the surrounding
CF. Each parametrization being characterized by its own multipolar crystal-field strengths
$S_{k}=\left(\frac{1}{2k+1}\sum\limits_{q}|B_{kq}|^{2}\right)^{1/2}$, for $k=2,4$ and $6$, yields a definite
second moment $\sigma$, which can be derived from the additivity relationship
$\sigma^{2}=\sum_{k}\sigma_{k}^{2}$ and the known asphericities $\langle \Psi||C^{(k)}||\Psi \rangle$
of the  central-ion eigenfunctions $\Psi$. The condition $\sigma=\sigma_{\rm exp}$ must be satisfied to ensure
the parametrization's correctness. However, our survey of literature indicates that there exists many other
well-fitted crystal-field parameter sets that do not obey this condition.
Therefore, such sets are erratic and non-physical, and should be re-examined or rejected.
Having $\sigma$ for several $(\geq3)$ eigenstates $|\Psi\rangle$ along with the relevant
$\langle \Psi||C^{(k)}||\Psi \rangle$ asphericities, one can estimate $\sigma_{k}$ and $S_{k}$, which are
well-founded experimentally. The above findings set up the parametrization process properly.
Lack of consistency between the second moments representing various parametrizations and the pertinent
second moments observed in experiments is presumably the main reason for deluge of formally accurate
but accidental and inequivalent parametrizations.
\end{abstract}
\noindent
{\it PACS}: 71.15.-m, 71.23.An, 71.70.Ch, 75.10.Dg \\
\section*{1. Introduction}
The review of the crystal-field (CF) Hamiltonian ${\cal H}_{\rm CF}=\sum_{k}\sum_{q}B_{kq}C_{q}^{(k)}$
parametrizations available in literature provides a vast amount of discouraging examples of their inconsistency and
divergence [1-6]. Only a slight part of them follows the standardization procedures [2-7].
Also the number of the inconsistent CF parametrizations seems to be alarming in the light of actual
multipolar structure of a typical crystal surroundings. A suspicion arises that the reason for such
widespread divergences among parametrizations is more fundamental than the choice of the reference system.

Although a complete mutual orthogonality of the component ${\cal H}_{\rm CF}$ multipoles and, in
consequence, their independent and separate contribution to the global second moment $\sigma$ have a crucial physical
meaning, this feature is not reflected in the conventional fitting procedures. The multipolar independence manifests
itself most fundamentally in the additivity of the relevant second moments $\sigma^{2}=\sum_{k}\sigma_{k}^{2}$
for any free-ion state CF splitting, where $\sigma_{k}$ is a partial (multipolar) second moment produced by the
$2^{k}$-pole ${\cal H}_{\rm CF}$ component [8-11] (section 2). Both $\sigma$ and $\sigma_{k}$ corresponding
to a given parametrization can be directly calculated. On the other hand, $\sigma$ and $\sigma_{k}$ can be derived
also from experimental data (section 4). There is a question - how do the two sets of values compare?

To answer this question we have analyzed thoroughly four splitting diagrams for the states
$^{4}$I$_{9/2}$, $^{4}$I$_{11/2}$,$^{4}$I$_{13/2}$ and $^{4}$I$_{15/2}$ of the $^{4}$I term of Nd$^{3+}$ ion doped
into single crystal of cubic yttrium oxide (Y$_{2}$O$_{3}$) [12] (section 4). Having obtained the second moments from
the splitting diagrams as well as the asphericities $A_{k}(|\Psi \rangle)=\langle \Psi||C^{(k)}||\Psi \rangle$ of the
considered eigenstates $|\Psi\rangle$ [13], and employing the partial second moments additivity, we have
estimated the physically well-founded $\sigma_{k}$ values and, in consequence, the relevant $S_{k}$. Unexpectedly we have
found that the proposed parametrization [12] leads to the second moments that are incompatible with $\sigma_{\rm exp}$
derived from the experiment. Therefore, the parametrization [12] has turned out to be an accidental and non-physical
one. It gives reasonable grounds for believing that there may exist different and inequivalent,
although formally accurate, ${\cal H}_{\rm CF}$ parametrizations leading to different second moments.
The erratic character of such fittings stems from a great number of well-fitted crystal-field parameter
(CFP) sets admissible when the relevant second moments are not correctly fixed.

This finding has prompted us to formulate the following criterion (the necessary condition) for the physical
fidelity of ${\cal H}_{\rm CF}$ parametrizations: \textbf{each and every correct CF parametrization has to yield
the second moment of any relevant splitting equal to the observed one}.
Disregard for this condition is presumably the main reason for an excessive number as well as inexplicable
discrepancies in parametrizations available in the literature. The point is that only the fitting procedures
which lead to $\sigma=\sigma_{\rm exp}$ can be treated as potentially correct ones.

The CFPs determination is inevitably connected with the accuracy of the initial approximation of the
involved eigenstates (section 3). The eigenfunctions applied in the presented paper have been obtained by
simultaneous diagonalization of the interaction matrix comprising Coulomb repulsion and spin-orbit coupling
using M. Reid f-shell programs [14]. Such type of functions (Table 2) are sufficiently accurate for most of the
states from lower parts of the rare earth ion spectra.

The effective character of CFPs in relation to the assumed basis of the initial eigenfunctions
forming the interaction matrix is shortly outlined in section 3. Possible effects of extending the applied
approximation of the initial eigenstates for highly excited states is also discussed.
Since these states might be strongly mixed by crystal-field [12,15, 16] the calculated asphericities $A_{k}$
can be markedly changed.

\section*{2. Basic formalism}
The effective CF Hamiltonian ${\cal H}_{\rm CF}=\sum_{k}\sum_{q}B_{kq}C_{q}^{(k)}$ for electrons with the quantum
number $l\leq 3$ is limited to the three or less $2^{k}$-poles for $k=2,4$ and $6$. These three terms are entirely
independent from each other as they transform themselves according to different irreducible representations
${\cal D}^{(k)}$ of the three -dimensional rotation group R$_{3}$.

Each individual $2^{k}$-pole contributes separately to the total splitting of any state. Undoubtedly, the relationship
between a partial (rank $k$) second moment $\sigma_{k}$, generated by a given $2^{k}$-pole, and the global second
moment $\sigma$, produced by the sum of the three multipoles, constitutes the adequate measure of the
multipolar contribution [8-11]:
\begin{equation}
\sigma^{2}|J\rangle=\frac{1}{2J+1}\sum\limits_{n} \left[ E_{n}-\overline{E}\left(|
J\rangle\right)\right]^{2}=\frac{1}{2J+1}\sum\limits_{k}S_{k}^{2} \left(\langle J||C^{(k)}||
J\rangle\right)^{2}=\sum\limits_{k}\sigma_{k}^{2},
\end{equation}
where the gravity center of the CF sublevels within the state $|J\rangle$ is given by
$\overline{E}(|J\rangle)=\frac{1}{2J+1}\sum\limits_{n}E_{n}$ with $E_{n}$ as the energy of $|n\rangle$ sublevel,
$S_{k}=\left(\frac{1}{2k+1}\sum\limits_{q}|B_{kq}|^{2}\right)^{1/2}$ stands for the conventional CF strength of the
$2^{k}$-pole ${\cal H}^{(k)}_{\rm CF}=\sum_{q}B_{kq}C_{q}^{(k)}$ [8-10,13,17], $\langle
J||C^{(k)}||J \rangle=A_{k}(|J\rangle)$ is the dimensionless scalar dependent only on the angular
distribution of the $|J\rangle$ state electron density and reflects its $2^{k}$-pole type asphericity
[13], whereas $|J\rangle$ denotes in a brief way any free-ion initial eigenstate in approximation
that preserves its quantum number $J$. Generally, when $J$ ceases to be a good quantum number the state
$|J\rangle$ is replaced by $|\Psi\rangle$ (section 1).

The CF strength $S_{k}$ is an invariant of the R$_{3}$ group and has clear physical
meaning due to its direct connection with $\sigma_{k}$ (Eq.(1)). The square of the global second moment
$\sigma^{2}$ is a simple sum of $\sigma_{k}^{2}$, and any compensation of the partial contributions is
excluded.
The parameter $S=\left(\sum_{k} S_{k}^{2}\right)^{1/2}$ or $S=\left(\frac{1}{3}\sum_{k} S_{k}^{2}\right)^{1/2}$
described earlier in the literature [8,17] as the global CF strength is another invariant of the R$_{3}$ group.
However, its physical meaning is rather obscure due to the scalar product $\sum_{k} S_{k}^{2}A_{k}^{2}$ in Eq.(1)
[13].

Although the additivity of the second moments $\sigma_{k}$ is crucial, this property seems to be underestimated
or even ignored in literure. The square of the global second moment $\sigma^{2}$ of any CF splitting can be always
written in the form of a positive definite sum:
\begin{equation}
\sigma^{2}(|J\rangle)=\frac{1}{2J+1}\left(S_{2}^{2}A_{2}^{2}(|J\rangle)+S_{4}^{2}A_{4}^{2}(|J\rangle)+
S_{6}^{2}A_{6}^{2}(|J\rangle)\right).
\end{equation}
Hence, knowing the energy spectra of $n$ states of a paramagnetic ion embedded in a given crystal-field, we have
at our disposal a set of $n$ linear equations as Eq.(2). Their algebraic nature admits only positive
solutions what markedly confines their solvability. In turn, based on the asphericities $A_{k}$ we can
determine all three $S_{k}$ or at least a relation between them, depending on the number of available equations.
All of these $S_{k}$ data well-founded experimentally should be built into the applied parametrization procedure.

The above linear equations (Eq.(2)) become simplified when some $S_{k}$ or $A_{k}$ vanish.
Vanishing of $S_{k}$ is governed by the symmetry and geometry of the central-ion surroundings.
All crystallographic point-symmetry groups, except those of the cubic system, admit the three $2^{k}$-poles for
$k=2,4$ and $6$. Only in the cubic system the quadrupole term is excluded, i.e. $S_{2}=0$. Nevertheless, $S_{k}$ may
vanish accidently due to compensation of the CFPs what results from a particular surroundings geometry.
In turn, vanishing of some $A_{k}$ depends on $J$ quantum number representing the resultant angular momentum
of the central ion eigenstate according to the triangle conditions for the three numbers $(J, J, k)$.
If $J\geq 3$, any $|J\rangle$ eigenstate is nominally characterized by three non-zero
$A_{k}$ parameters for $k=2,4,6$. Nevertheless, it may happen that due to the mixing of the states some $A_{k}$
can compensate (or enhance) themselves to a different degree (Table 2).
For $J=2$ or $5/2$ only $A_{2}$ and $A_{4}$ differ from zero leading to Eqs (2) with at most two unknowns
$S_{2}^{2}$ and $S_{4}^{2}$. For $J=1$ or $3/2$ solely the quadrupole term is effective, i.e.
only $A_{2}\neq 0$. The total splitting $\Delta E$ of the states in the last case allows us to estimate $S_{2}$ directly
knowing accurately the relevant $A_{2}$. As an example, for any state $|J=3/2\rangle$ we obtain
$\sigma^{2}=\frac{1}{4}\left(\Delta E\right)^{2}=\frac{1}{4}S_{2}^{2}A_{2}^{2}$, and hence
\begin{equation}
S_{2}=\frac{\Delta E}{A_{2}}.
\end{equation}
Since $\Delta E$ depends exclusively on $S_{2}$, the pertinent individual $B_{2q}$ CFPs cannot be determined
in this way.

\section*{3. The effective character of crystal-field parameters}
The actual CFPs should characterize only the multipolar structure of the central-ion surroundings in crystals.
The point is, however, that they are always accessible in products with the corresponding multipolar components
$\langle\Psi||C^{(k)}||\Psi \rangle$ of electron density distribution in the central-ion open shell.
Thus, CFPs that precisely reproduce the actual energies quite naturally depend on the $\Psi$ eigenfunction basis.
This is why CFPs may take different values according to the basis of the assumed eigenfunctions $\Psi$.
Let us emphasize that only the accurate eigenfunctions provide the right CFP set which has clear physical meaning
and a universal character as it is irrespective of a central-ion.
In fact, one should realize that we always deal with more or less "dressed" CFPs, whereas our aim are
the "bare" ones. Obviously, even the best fitted CFP set obtained for not particularly accurate central-ion
eigenfunctions cannot be recognized as a definitive solution.

CFPs obtained in fitting procedures are only roughly approximated for Russell-Saunders (RS) eigenfunctions, better
after simultaneously taking into account Coulomb repulsion and spin-orbit coupling, and the best approximated
after simultaneous diagonalization of Coulomb repulsion, spin-orbit coupling, and CF interactions.
Moreover, within a given electron configuration, contributions of the above mentioned interactions to the eigenstates
of different energy may differ substantially.

In the energy spectra of tripositive rare earth ions the low lying states are quite well approximated
considering simultaneously Coulomb repulsion and spin-orbit coupling. They are somewhat worse approximated
including the spin-orbit coupling in a perturbational way ($J$-mixing approach) as well as even for the pure RS coupling.
Nevertheless, for higher excited states all these approximations fail. Due to the higher density of these states,
i.e. proximity of their energies, the CF mixing causes noticeable or even considerable changes in their
compositions and energies [12,15].

Applying uncritically some of the above simplified approximations for the whole basis of $4f^{N}$ states
may hinder appropriate parametrization. As a matter of fact, this approach might be adequate for a
part of lower levels, whereas for the higher ones not necessarily.
The incompatibility between $S_{2}$ estimated from the CF splitting of the  $^{4}|F_{3/2}\rangle$ state
of Nd$^{3+}$:Y$_{2}$O$_{3}$ [12] (Eq.(3)) and its value resulting from the second moment analysis (section 4),
may serve as an example of such disparity.
The splitting $\Delta E$ found experimentally amounts to $196\;cm^{-1}$. Using the approximate $A_{2}$ value for
the state $^{4}|F_{3/2}\rangle$ equal to $0.3561$ (Table 2) one gets $S_{2}=550\;cm^{-1}$, which reaches only
$56\%$ of the needed magnitude (section 4). It results from an overestimated $A_{2}$ value, which in fact should
be close to $0.200$.

Another example is an unexpected dominance of the CF splitting of $|^{3}$F$_{2}\rangle$ state (at $5200\;cm^{-1}$)
over that of $|^{3}$P$_{2}\rangle$ state (at $22750\;cm^{-1}$) of Pr$^{3+}$ ion LaF$_{3}$ matrix [15].
It reflects well the scale of changes in their eigenfunctions and asphericities (mainly for the $|^{3}$P$_{2}\rangle$
state).
For the $|^{3}$F$_{2}\rangle$ state we have $\Delta E=143\;cm^{-1}$ and $\sigma=55.1\;cm^{-1}$, whereas for the
$|^{3}$P$_{2}\rangle$ state $\Delta E=128\;cm^{-1}$ and $\sigma=45.2\;cm^{-1}$ [15]. However, $|A_{k}|$ of the state
$|^{3}$F$_{2}\rangle$ calculated within the standard approximation are less than those for the
$|^{3}$P$_{2}\rangle$ state and amount to: $|A_{2}|=0.3400$, $|A_{4}|=0.0909$ versus $|A_{2}|=0.8563$, $|A_{4}|=0.1112$.

Let us remind also that the conventional CFPs contain intentionally the factor $\langle r^{k}\rangle$. This
quantities -- the mean $k$ powers of the radii of the magnetic electrons [18] -- concerns the central-ion itself.
Nevertheless, since the $\langle r^{k}\rangle$ are only simple scaling factors, it is easy to renormalize the
conventional CFPs to the ones independent of a central-ion.

\section*{4. Results}

The method of adjustment of the second moments presented below has been based on the CF splitting diagrams
of the four lowest states originating from the $^{4}$I term of $4f^{3}$ configuration of Nd$^{3+}$ ion in
Nd$^{3+}$:Y$_{2}$O$_{3}$ [12]. The Nd$^{3+}$ ions occupy the $C_{2}$ lattice sites and all the free-ion states
are split into Kramers doublets.
The data come from the optical absorption and fluorescence spectra reported by Chang et al [12] (Table 1). The
relevant asphericities $A_{k}$ of the involved states calculated for the eigenfunctions obtained using
M.Reid f-shell programs [14] and the free-ion data reported by Carnall et al [16] are compiled in Table 2.

The additivity of the partial second moments leads to the following set of linear equations for
$S_{k}^{2}$ (Table 1):
\begin{eqnarray}
&&\frac{1}{10}\left[S_{2}^{2}A_{2}^{2}\left(^{4}I_{9/2}\right)+ S_{4}^{2}A_{4}^{2}\left(^{4}I_{9/2}\right)+
                    S_{6}^{2}A_{6}^{2}\left(^{4}I_{9/2}\right)\right]=\sigma^{2}\left(^{4}I_{9/2}\right) \nonumber \\
&&\frac{1}{12}\left[S_{2}^{2}A_{2}^{2}\left(^{4}I_{11/2}\right)+ S_{4}^{2}A_{4}^{2}\left(^{4}I_{11/2}\right)+
                    S_{6}^{2}A_{6}^{2}\left(^{4}I_{11/2}\right)\right]=\sigma^{2}\left(^{4}I_{11/2}\right)\nonumber
                    \\
&&\frac{1}{14}\left[S_{2}^{2}A_{2}^{2}\left(^{4}I_{13/2}\right)+ S_{4}^{2}A_{4}^{2}\left(^{4}I_{13/2}\right)+
                    S_{6}^{2}A_{6}^{2}\left(^{4}I_{13/2}\right)\right]=\sigma^{2}\left(^{4}I_{13/2}\right)\nonumber \\
&&\frac{1}{16}\left[S_{2}^{2}A_{2}^{2}\left(^{4}I_{15/2}\right)+ S_{4}^{2}A_{4}^{2}\left(^{4}I_{15/2}\right)+
                    S_{6}^{2}A_{6}^{2}\left(^{4}I_{15/2}\right)\right]=\sigma^{2}\left(^{4}I_{15/2}\right).
\end{eqnarray}
The RS labels in the parentheses are assigned according to the maximal component in the free-ion
eigenvector of the state. Substituting the values of $\sigma^{2}\left(^{4}I_{J}\right)$ from Table 1 and
$A_{k}\left(^{4}I_{J}\right)$ from Table 2 into Eqs (4) we obtain in $[(cm^{-1})^{2}]$:
\begin{eqnarray}
&&\frac{1}{10}\left(0.2264S_{2}^{2}+ 0.2249S_{4}^{2}+ 1.1606S_{6}^{2}\right)=60238 \nonumber \\
&&\frac{1}{12}\left(0.2478S_{2}^{2}+ 0.1518S_{4}^{2}+ 0.1138S_{6}^{2}\right)=33507 \nonumber \\
&&\frac{1}{14}\left(0.3153S_{2}^{2}+ 0.2209S_{4}^{2}+ 0.3827S_{6}^{2}\right)=40373 \nonumber \\
&&\frac{1}{16}\left(0.4339S_{2}^{2}+ 0.4742S_{4}^{2}+ 3.1560S_{6}^{2}\right)=83211.
\end{eqnarray}
We have found some acceptable solutions centered at the values: $S_{2}^{2}=950000\;(cm^{-1})^{2}$,
$S_{4}^{2}=950000\;(cm^{-1})^{2}$ and $S_{6}^{2}=150000\;( cm^{-1})^{2}$ which can be recognized as reliable.
This non-optimized solution satisfies Eqs (5) with \emph{rms} deviation $368\;(cm^{-1})^{2}$, i.e. within the
accuracy of ca. $99\%$.

On the other hand the CFPs for Nd$^{3+}$:Y$_{2}$O$_{3}$ given in Table VI [12] yield the
following $S_{k}^{2}$ in $[(cm^{-1})^{2}]$: $S_{2}^{2}=278472$, $S_{4}^{2}=756148$, and $S_{6}^{2}=76576$,
which are smaller than those calculated from Eqs (5), particularly $S_{2}^{2}$.
Substituting now the asphericities $A_{k}$ of the involved states (Table 2) into Eq.(1) we
obtain the following $\sigma^{2}$ in $[(cm^{-1})^{2}]$: 33094, 16548, 20926
and 46250 for the corresponding split states $^{4}I_{9/2}$, $^{4}I_{11/2}$, $^{4}I_{13/2}$ and $^{4}I_{15/2}$.
They reach only a half of the observed $\sigma^{2}$ (Table 1). So, the proposed parametrization does not guarantee
the required second moments of the observed splittings, and in consequence is devoid of physical meaning. Its
pretty high accuracy (\emph{rms} $=7.0\;cm^{-1})$ is misleading as it refers to an accidental and improper
$\sigma$. The correct parametrization has to provide any other CFP set that should ensure both the same energy
of the CF sublevels and the correct second moments. Such solution is always provided by each equation
$\sigma^{2}(|J\rangle)=\sum_{k}S_{k}^{2}A_{k}{^2}(|J\rangle)$.

\section*{5. Discussion}
A typical example of CF parametrization analyzed in section 4 reveals that the set of CFPs derived from
spectroscopic data in an advanced fitting procedure (Table VI in [12]) yields only about 50$\%$ of the observed
squares of the global and partial second moments of the splittings.
It means that the physically correct parametrization, which gives the appropriate second moments,
turns out to be correlated with a quite different $\sigma=\sigma_{\rm exp}$.
It points to the important conclusion that many coincidental ${\cal H}_{\rm CF}$ parametrizations can be found for
any considered splitting if there are no restrictions imposed on the relevant second moments.
Such possible accidental parametrizations can differ in the following ways:
\begin{itemize}
    \item [(i)] in the global second moment $\sigma$,
    \item [(ii)]in a different partition $\sigma^{2}=\sum_{k}\sigma_{k}^{2}$ at the same $\sigma$,
    \item [(iii)] in a different mutual spatial orientation of the component multipoles [19] at the same
    $\sigma$ and the same $\sigma_{k}$ [19].
\end{itemize}
Hence, all the CFP sets fulfilling conditions (i)--(iii) are, to a different extent, inequivalent
and therefore false.
The correct and equivalent parametrizations should yield $\sigma=\sigma_{\rm exp}$, the identical proper partition
$\sigma^{2}=\sum_{k}\sigma_{k}^{2}$ and the same mutual spatial orientation of the component multipoles.
All the remaining parametrizations not necessarily correspond to certain local \emph{rms} minima occurring in the
fittings, but can be merely accidental and non-physical ones which should be verified.
Due to their diversified $S_{k}$ (the rotational invariants) or various spatial orientations of the
component multipoles they do not follow the standardization procedures based on the reference frame rotations
[2,3,4,7,19]. While verifying inequivalent parametrizations (concerning the same paramagnetic ion in a lattice
system) one should a priori impose the known $\sigma_{\rm exp}$. It should reduce the parametrizations to the correct
and equivalent ones, differing at most in the relevant reference frame.

Disregard for the second moments in ${\cal H}_{\rm CF}$ parametrization explains well the
inflation of various CFP sets obtained in fittings. There are no other convincing reasons for such deluge of solutions.
After all, the actual multipolar structure of the parametrized CF unknown during the fitting process
is rather strictly defined. A question arises -- what is the probability of a fictitious partition
$\sigma^{2}=\sum_{k}\sigma_{k}^{2}$ in the fitting process when $\sigma=\sigma_{\rm exp}$ is fixed a priori?
Such probability should not be high since each partition corresponds to characteristic proportions of the component
multipoles in ${\cal H}_{\rm CF}$. So, a coincidence of the inequivalent parametrizations seems to be an exceptional
case. The orientation of the nominal reference frames (unknown by definition) [4,6] during searching for the proper
partition is inessential.
Let us notice that identities of both $\sigma$ and $\sigma_{k}$ correspondingly for different coincidental
parametrizations are the necessary conditions for their equivalence and fidelity.
However, some equivalent CFP sets can be false if their identical $\sigma$ and $\sigma_{k}$ are not equal to the
respective actual (observed) values.

In order to derive partial $\sigma_{k}$ and hence $S_{k}$ from the relevant linear equations (Eqs (4)) and
known $\sigma_{\rm exp}$, we still require the asphericities $A_{k}$ of the involved states. The $A_{k}$ accuracy
depends on how adequate are the assumed eigenfunctions and hence determines the precision of both $\sigma_{k}$
and $S_{k}$.
Therefore, it is recommended to use the CF splitting data for low-energy states of well-defined eigenfunctions.
On the other hand, in some cases even the simple RS functions can turn out to be sufficiently good (Table 2).
For example, the eigenfunctions of the states $|^{3}H_{5}\rangle$, $|^{3}F_{2}\rangle$ and $|^{3}P_{1}\rangle$ for
configurations $4f^{2}$ or $4f^{12}$ in Pr$^{3+}$ or Tm$^{3+}$ ions remain single-component eigenvectors (within the
assumed approximation) since in these configurations there are no other states with the same $J$.

For higher excited states, to approximate the eigenfunctions sufficiently well, the full simultaneous diagonalization
of the complete interaction matrix including the CF Hamiltonian is needed due to the importance of CF effect.
But in such case a more sophisticated self-consistent numerical approach to the problem should be applied.

The CF analysis carried out in [12] was based on a Hamiltonian of $C_{2}$ point group symmetry including
$J$-mixing effects. The program by M. Reid [14] employs a simultaneous diagonalization of the interaction matrix
including both Coulomb repulsion and spin-orbit coupling. The discrepancies in the resulting $A_{k}$'s are negligible.
Actually they are close to the asphericities of the pure RS states (Table 2).

The linear equations (see Eq. (2)) relating $S_{k}$ and $A_{k}$ can be employed alternatively. One can find $S_{k}$
using $A_{k}$, or vice versa, obtain $A_{k}$ from $S_{k}$. Having the global second moments $\sigma$
for several states, e.g. Eqs (4), we may reconcile these quantities as it has been shown in section 3 for the
$|^{4}F_{3/2}\rangle$ and four lowest states of the Nd$^{3+}$:Y$_{2}$O$_{3}$ [12].

\section*{6. Conclusions}

Based on Nd$^{3+}$:Y$_{2}$O$_{3}$ spectroscopic data and the related CF analysis [12] we have shown that
formally correct CF parametrization does not lead to the experimental second moments
of the relevant CF splittings. There are reasonable grounds for believing that it is not an exceptional case
among available CF parametrizations. Presumably, it can be also the main reason for their inconsistencies.
Thus, the fundamental criterion to be fulfilled by the physically founded and correct parametrization has to
yield the second moment of any pertinent splitting equal to the observed one. This necessary condition sounds trivially
only because there exists a widespread believe that it is satisfied automatically in used fitting procedures.
In order to verify a given parametrization it is enough to know the asphericities
$A_{k}(\Psi)=\langle \Psi||C^{(k)}||\Psi \rangle$, $k=2,4$ and $6$, for the initial central-ion eigenfunctions
$\Psi$ along with the corresponding $\sigma_{\rm exp}$ and check the relationship
$\sigma_{\rm exp}^{2}=\sum_{k}S_{k}^{2}A_{k}^{2}(\Psi)$. To avoid false, non-physical parametrizations this condition
should be explicitly built in the fitting programs.

\section*{Acknowledgments}
We are indebted to Z. Gajek for providing us with the free-ion eigenstates calculated using
M.Reid f-shell programs.




\renewcommand{\baselinestretch}{1}
\clearpage
\begin{small}
\begin{table*}[htbp]
\begin{center}
\caption{Experimental crystal-field splitting data for $^{4}$I states of Nd$^{3+}$:Y$_{2}$O$_{3}$ [12]}

\vspace*{0.6cm}
\begin{tabular}{lcccc}
\hline
State $^{\ast}$       & Centroid parameter       & Number of     &  Total splitting                   & Square of the second  \\
$\left[SL\right]$J    &  $\left[cm^{-1}\right]$  &  CF doublets  & $\Delta E$ $\left[ cm^{-1}\right]$ & moment $\sigma^{2}$   $\left[ cm^{-1}\right]^{2}$\\
\hline
$^{4}$I$_{9/2}$   & 314 & 5   & 643  & 60238 \\
\hline
$^{4}$I$_{11/2}$   & 2171 & 6   & 462  & 33507 \\
\hline
$^{4}$I$_{13/2}$   & 4133 & 7   & 515  & 40373 \\
\hline
$^{4}$I$_{15/2}$   & 6160 & 8   & 770  & 83211 \\
\hline \multicolumn{5}{c}{$^{\ast}$  {\small Russell-Saunders labels according to the maximal component in the free-ion eigenvector of the state}}\\
\end{tabular}
\end{center}
\end{table*}
\end{small}

 \clearpage

\renewcommand{\baselinestretch}{1}
\clearpage
\begin{small}
\begin{table*}[htbp]
\begin{center}
\caption{Asphericities of chosen Nd$^{3+}$ free-ion eigenstates (the values for the upper component of the state
are given in parentheses)}
\begin{tabular}{llccc}
\hline No     &    $^{\ast}$Eigenstate         $|\Psi \rangle$         & \multicolumn{3}{c}{Multipolar asphericity} \\
              &    \emph{\textbf{Energy}} $\left[cm^{-1}\right]$     & $A_{2}(\Psi)$   &  $A_{4}(\Psi)$  & $A_{6}(\Psi)$             \\
              &                       & $\langle \Psi||C^{(2)}||\Psi \rangle$
                                      & $\langle \Psi||C^{(4)}||\Psi \rangle$
                                      & $\langle \Psi||C^{(6)}||\Psi \rangle$       \\
\hline
1 & 0.9844$|^{4}$I$_{9/2}\rangle$-0.1647$|^{2}$H(2)$_{9/2}\rangle$+0.0566$|^{2}$H(1)$_{9/2}\rangle$ &  -0.4758 & -0.4742 & -1.0773  \\
            &   \hspace*{5cm} + 2 {\it components}              &      &     &    \\
            & \emph{\textbf{0}}                 &      &     &    \\
         &\hspace*{1.1cm} $|^{4}$I$_{9/2}\rangle$ &  (-0.4954) & (-0.4904) & (-1.1085) \\
\hline
2 & 0.9947$|^{4}$I$_{11/2}\rangle$-0.0949$|^{2}$H(2)$_{11/2}\rangle$+0.0361$|^{2}$H(1)$_{11/2}\rangle$ &  -0.4978 & -0.3896 & -0.3374  \\
            &   \hspace*{5cm} + 1 {\it component}              &      &     &    \\
            & \emph{\textbf{1862}}                 &      &     &    \\
         &\hspace*{1.1cm} $|^{4}$I$_{11/2}\rangle$ &  (-0.5045) & (-0.3935) & (-0.3399) \\
\hline
3 & 0.9979$|^{4}$I$_{13/2}\rangle$+0.0609$|^{2}$K$_{13/2}\rangle$-0.0231$|^{2}$I$_{13/2}\rangle$ &  -0.5615 & -0.4700 & -0.6186  \\
            &                 &      &     &    \\
            & \emph{\textbf{3845}}                 &      &     &    \\
         &\hspace*{1.1cm} $|^{4}$I$_{13/2}\rangle$ &  (-0.5569) & (-0.4691) & (-0.6217) \\
\hline
4 & 0.9938$|^{4}$I$_{15/2}\rangle$+0.1110$|^{2}$K$_{15/2}\rangle$&  -0.6587 & -0.6886 & -1.7765  \\
            &                 &      &     &    \\
            & \emph{\textbf{5907}}                 &      &     &    \\
         &\hspace*{1.1cm} $|^{4}$I$_{15/2}\rangle$ &  (-0.6438) & (-0.6850) & (-1.7999) \\
\hline
5 & 0.9698$|^{4}$F$_{3/2}\rangle$+0.2232$|^{2}$D(1)$_{3/2}\rangle$-0.0614$|^{2}$D(2)$_{3/2}\rangle$ &  0.3561 & 0 & 0  \\
            &   \hspace*{5cm} + 3 {\it components}              &      &     &    \\
            & \emph{\textbf{11381}}                 &      &     &    \\
         &\hspace*{1.1cm} $|^{4}$F$_{3/2}\rangle$ &  (0.3578) & (0) & (0) \\
\hline
6 & 0.9879$|^{4}$F$_{5/2}\rangle$+0.1468$|^{2}$D(1)$_{5/2}\rangle$-0.0321$|^{2}$F(2)$_{5/2}\rangle$ &  0.3277 & 0.1843 & 0  \\
            &   \hspace*{5cm} + 3 {\it components}              &      &     &    \\
            & \emph{\textbf{12420}}                 &      &     &    \\
         &\hspace*{1.1cm} $|^{4}$F$_{5/2}\rangle$ &  (0.3220) & (0.1890) & (0) \\
\hline
7 & 0.9648$|^{4}$F$_{7/2}\rangle$+0.2004$|^{2}$G(1)$_{7/2}\rangle$-0.1582$|^{2}$G(2)$_{7/2}\rangle$ &  0.4650 & 0.0076 & -0.4233  \\
            &   \hspace*{5cm} + 3 {\it components}              &      &     &    \\
            & \emph{\textbf{13383}}                 &      &     &    \\
         &\hspace*{1.1cm} $|^{4}$F$_{7/2}\rangle$ &  (0.4601) & (0.0537) & (-0.4552) \\
\hline
8 & -0.8670$|^{4}$F$_{9/2}\rangle$-0.4420$|^{2}$H(2)$_{9/2}\rangle$+0.1526$|^{2}$H(1)$_{9/2}\rangle$ &  0.5748 & -0.3050 & 0.1040  \\
            &   \hspace*{5cm} + 4 {\it components}              &      &     &    \\
            & \emph{\textbf{14652}}                 &      &     &    \\
         &\hspace*{1.1cm} $|^{4}$F$_{9/2}\rangle$ &  (0.7136) & (-0.4051) & (0.1799) \\
\hline
9 & 0.7205$|^{2}$P$_{3/2}\rangle$+0.6371$|^{2}$D(1)$_{3/2}\rangle$-0.1869$|^{2}$D(2)$_{3/2}\rangle$ &  0.2818 & 0 & 0  \\
            &   \hspace*{5cm} + 3 {\it components}              &      &     &    \\
            & \emph{\textbf{26179}}                 &      &     &    \\
         &\hspace*{1.1cm} $|^{2}$P$_{3/2}\rangle$ &  (0.2981) & (0) & (0) \\
\hline
\multicolumn{5}{l}{$^{\ast}${\small The eigenfunctions and eigenvalues have been calculating using M. Reid f-shell programs [14] and free-ion data}}\\
\multicolumn{5}{l}{{\small  reported by Carnall et al [16]. Only the first three main components of the states are given explicitly.}}\\
\multicolumn{5}{l}{{\small }}\\
\end{tabular}
\end{center}
\end{table*}
\end{small}


\begin{thebibliography}{99}
\renewcommand{\baselinestretch}{1}

\bibitem{1}    C. A. Morrison, R. P. Leavitt,
                Spectroscopic properties of triply ionized lanthanides in transparent host crystals
                in: K. A. Gschneidner, Jr. and L. Eyring (eds),
                Handbook of the Physics and Chemistry of Rare Earths,
                (North-Holland, Amsterdam, 1982).
\bibitem{2}    C. Rudowicz and R. Bramley, J. Chem. Phys. {\bf 83}, 5192 (1985).
\bibitem{3}    C. Rudowicz, J. Chem. Phys. {\bf 84}, 5045 (1986).
\bibitem{4}    C. Rudowicz and J. Qin, J. Lumin. {\bf 110}, 39 (2004).
\bibitem{5}    C. Rudowicz and J. Qin, J. Alloys and Compounds {\bf 385}, 238 (2004).
\bibitem{6}    C. Rudowicz and P. Gnutek, Physica B {\bf 405}, 113 (2010).
\bibitem{7}    G.W. Burdick and M.F. Reid, Mol. Phys. {\bf 102}, 1141 (2004).
\bibitem{8}    R. P. Leavitt, J. Chem. Phys. {\bf 77}, 1661 (1982).
\bibitem{9}    Y.Y. Yeung, Invariants and moments in: Crystal Field Handbook, edited by D.J. Newman and B. Ng
               (Cambridge Univrsity Press, Cambridge, MA, 2000).
\bibitem{10}   C. Rudowicz and J. Qin, Phys. Rev. {\bf B67}, 174420 (2003).
\bibitem{11}    J. Mulak and M. Mulak, J. Phys. A: Math. Theor. {\bf 40}, 1 (2007).
\bibitem{12}    N. C. Chang, J. B. Gruber, R. P. Leavitt and C. A.Morrison J. Chem. Phys. {\bf 76}, 3877 (1982).
\bibitem{13}   J. Mulak and M. Mulak, Phys. Stat. Solidi B {\bf 245}, 1156 (2008).
\bibitem{14}   M. Reid {\it f-shell programs} (2010) (by courtesy of Z. Gajek).
\bibitem{15}   W.T. Carnall and H. Crosswhite, J. Less-Common Metals {\bf 93}, 127 (1983).
\bibitem{16}   W.T. Carnall, G.L.Goodman, K. Rajnak and R.S. Rana, J.Chem. Phys. {\bf 90}, 3443 (1989).
\bibitem{17}   F. Auzel and O. L. Malta, J. Physique {\bf 44}, 201 (1983).
\bibitem{18}   A. J. Freeman and R. E. Watson, Phys. Rev. {\bf 127}, 2058 (1962).
\bibitem{19}   J. Mulak and M. Mulak, J. Phys. A: Math. Theor. {\bf 38}, 6081 (2005).
\end{thebibliography}
\end{document}